\documentclass[twocolumn,showpacs,preprintnumbers,amsmath,amssymb]{revtex4}
\usepackage{graphicx}
\usepackage{epstopdf}
\usepackage{dcolumn}
\usepackage{bm}
\usepackage{multirow}

\begin{document}

\renewcommand{\baselinestretch}{1.1} 


\title{Analysis of major failures in Europe's power grid}

\author{Mart\'i Rosas-Casals$^{1,2}$ and Ricard  Sol\'e$^{2,3}$ }
\affiliation{
$^1$ C$\grave{a}$tedra UNESCO de Sostenibilitat, Universitat Polit$\grave{e}$cnica de Catalunya (UPC), EUETIT-Campus Terrassa, Edif. TR4, C. Colom, 1, 08222 Barcelona, Spain\\ 
$^2$ ICREA-Complex Systems Lab, Universitat Pompeu Fabra - PRBB, Dr. Aiguader 88, 08003 Barcelona, Spain\\
$^3$ Santa Fe Institute, 1399 Hyde Park Road, New Mexico 87501, USA}

\begin{abstract}
Power grids are prone to failure. Time series of reliability measures such as total power loss or energy not supplied can give significant account of the underlying dynamical behavior of these systems, specially when the resulting probability distributions present remarkable features such as an algebraic tail, for example. In this paper, seven years (from 2002 to 2008) of Europe's transport of electricity network failure events have been analyzed and the best fit for this empirical data probability distribution is presented. With the actual span of available data and although there exists a moderate support for the power law model, the relatively small amount of events contained in the function's tail suggests that other causal factors might be significantly ruling the system's dynamics.
\end{abstract}





\maketitle


\section{Introduction}
There has been in recent years an increasing awareness about infrastructure networks security and reliability \cite{Wong09, Lewis06, Perrow07}. Modern society's functional capacity relays on an optimal operation of infrastructure and information networks such as roads, railways, gas and oil pipes or Internet. Particularly vital, and at the same time quite prone to failure, are electric power transmission networks. These are extremely complex engineered systems, composed of multiple and interconnected elements, whose reliability depends both on each component's behavior and, at the same time, on the many different dynamical interactions that span over and rule the overall connectivity of the system.

Although it is not always the case, a malfunction of a power transmission system shows usually itself as a blackout. This is a  direct consequence of a cascading failure involving several of its composing and linking elements. This fact turns the study of the details of failures in power transmission networks from a traditional engineering point of view a hard task, if not an impossible one most of the times. In order to reduce the inherent complexity of this detailed approach, some new ways have been proposed in recent years. One of them is that of ignoring the details of particular failures and to focus on the study of global behaviors and dynamics of time series with approximate global models. Concepts such as criticality and self-organization have been applied to characterize blackout data, suggesting that the frequency of large blackouts is governed by non trivial distribution functions such as power laws and, consequently, that power systems are designed and operated near a critical point. (For a comprehensive review on this approach, see Ref.\cite{Dobson2007} and references therein).

This paper analyses for the first time, and as far as we know, the statistics of major electric transmission network events in the European power grid from this aforementioned complex systems approach. Following essentially the statistical analysis presented in Ref.\cite{Clauset2007}, we estimate the basic parameters of the power-law model, then calculate the goodness-of-fit between the data and the power law and finally 
we compare the power law with alternative hypotheses via a likelihood ratio test. The paper is organized as follows. In section II European major events data is presented and explained. In section III blackout data is analyzed. Finally, section IV summarizes our main results.

\begin{figure*}[htbp]
\includegraphics[clip=true, width=.8\textwidth]{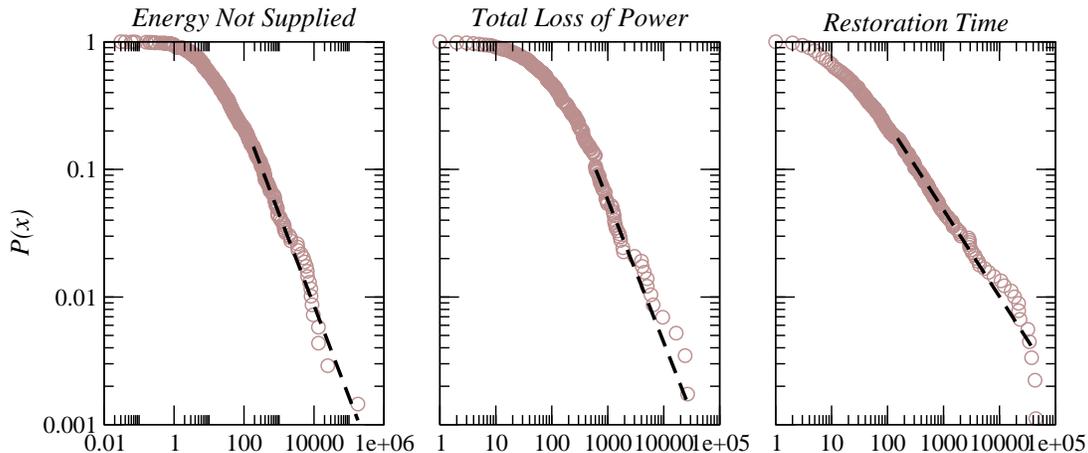}
\caption{Cumulative distribution functions $P(x)$ and their maximum likelihood power law fits for the UCTE reliability measures energy not supplied, total loss of power and restoration time.}
\label{fig:UCTEfittings}
\end{figure*}


\section{UCTE major events data}
European power network reliability data can be found in the Union for the Co-ordination of Transmission of Electricity (UCTE) web page, publicly available from 2002 onwards in monthly statistics format \cite{UCTEmonth}. The UCTE is the association of Transmission System Operators (TSOs) in continental Europe and manages data from 24 different European countries. Due to the complexity of events, sometimes involving more than one TSO, types of interruptions in the network and short time given to gather this information,
UCTE major events data is somehow limited in its scope and does not provide a fully detailed account of some events. It is, nonetheless,the best documented source that has been found. For each major event, it summarizes the following information:

\begin{itemize}
  \item \textbf{Country}.
  \item \textbf{Substations involved}.
	\item \textbf{Reason (R)}. Broadly classified into four groups: (1) overloads (also calculated brakes), (2) failures (false operation, failure in protection device or other element),	(3) external (outside impacts and very exceptional weather and natural conditions) and (4) other or unknown reasons.
	\item \textbf{Energy Not Supplied (ENS)}. Measured in MWh, as loss of energy from the consumption side.
	\item \textbf{Total Loss of Power (TLP)}. Measured in MW, as loss of production from the generation side.
	\item \textbf{Restoration Time (RT)}. Measured in minutes. Note that since ENS and TLP are measured from different sides, RT can not be assumed as the ratio of ENS over TLP. It can be considered, therefore, an independent reliability measure. 
	\item \textbf{Equivalent Time of Interruption (ETI)}. 
	Defined as the duration of an interruption in minutes multiplied by the energy not supplied divided by the consumption for the last 12 months. Defined in this way, the ETI allows a direct comparison between TSOs in terms of interruptions that occurred during a year. 
\end{itemize}

From 2002 to 2008, both years inclusive, 908 major events have been noticed. Due to the complexity of events some entries have zero value in one or more of their categories. While these zeroed values have not been considered, the rest of numerical values are effective measures of major events occurred in the UCTE power grid and, consequently, they have all been included in order to develop the analysis presented in this paper.

\section{Probability distribution analysis}

\begin{table*}
\boldmath
\renewcommand{\tabcolsep}{2mm}
\renewcommand{\arraystretch}{1.2} 
\begin{center} 
\begin{tabular}{ c c c c c| c c c c|c|}
\cline{6-9}
 & & & & & \multicolumn{4}{c|}{\textbf{Maximum likelihood}} \\ 
\cline{1-10}
\multicolumn{1}{|c|} {\textbf{Data set}} &  $n$ & $\left\langle{x}\right\rangle$ & $\sigma$ & $x_{max}$ & $\hat{x}_{min}$ & $\hat{\alpha}$ & $n_{tail}$ & $p$ & \textbf{Support for PL} \\
\cline{1-10}
\multicolumn{1}{|c|}{\textbf{ENS}} 	&  690 & 552 & 7004 & 180000 &  185$\pm$72 & 1.7$\pm$0.1 &  104$\pm$120 &  0.24 & Moderate\\ 
\multicolumn{1}{|c|}{\textbf{TLP}} 	&  576 & 400 & 1790 & 26746  & 615$\pm$244	 & 2.1$\pm$0.2 &  57$\pm$96 &  0.36 & Moderate\\ 
\multicolumn{1}{|c|}{\textbf{RT}} 	&  897 & 510 & 3328 & 44640  & 150$\pm$68	& 1.69$\pm$0.07 &  157$\pm$115 &  0.73 & Ok \\ 
\cline{1-10}                       
\end{tabular}
\end{center}
\unboldmath
\caption{UCTE major events generic statistics and power law fits. For each measure we give the number of occurrences $n$, mean $\left\langle{x}\right\rangle$, standard deviation $\sigma$, maximum observed occurrence $x_{max}$, lower bound to the power law behavior $\hat{x}_{min}$, scaling parameter value $\hat{\alpha}$, occurrences in the power law tail $n_{tail}$ and p value  $p$. The last column indicates the support for whether the observed data is well approximated by a power-law distribution. Estimated uncertainties for $\hat{x}_{min}$, $\hat{\alpha}$ and $n_{tail}$ are also shown.}
\label{tab:UCTEfittings}
\end{table*}

 The study of the statistics and dynamics of series of events with approximate global models has been one of the most popular topics in the last twenty years, specially within the interdisciplinary study of complex systems. Probability distribution functions with a heavy tailed dependence in terms of event or object sizes seem to be ubiquitous statistical features of self-organized natural and social complex systems \cite{Sornette2007}. The appearance of algebraic distributions, specially power laws, is often thought to be the signature of hierarchy, robustness, criticality and universal subjacent mechanisms \cite{Bak}. Electric power transmission networks have not escaped this captivation for power laws quest. Time series of usual measures of blackout size like energy unserved, power loss or number of customers affected, have been shown to be algebraically distributed in North America \cite{Carreras2002}, Sweden \cite{Holmgren2006}, Norway \cite{Bakke2006}, New Zealand \cite{Jordan2006} and China \cite{Weng2006}. This apparent ubiquitous evidence have led to believe and try to demonstrate that power systems (a) tend to self-organize near a critical point and (b) that there may be some universality ruling the inner depths of these systems.

In spite of this \textit{evidence}, most of the aforementioned literature relay on poorly performed statistical analysis and results can not be trusted. In some cases methodologies are not clearly explained (i.e., Ref.\cite{Bakke2006}) or simple visual inspection can clearly dismiss the analysis performed to rule in the power law hypothesis (i.e., Ref.\cite{Holmgren2006} and Ref.\cite{Weng2006}). In other cases, proper usage of statistical tools have given new results that limit the scope of the original analysis (i.e., Ref.\cite{Carreras2002} is dismissed as insufficiently substantiated in Ref.\cite{NewmanPareto} and reanalyzed in Ref.\cite{Clauset2007}, finding moderate support for the power law hypothesis and even some  for an exponential distribution).

In this section we analyze the probability distributions of three malfunction measures of the European power grid: energy not supplied (ENS), total loss of power (TLP) and restoration time (RT). The results are summarized in Table \ref{tab:UCTEfittings} and shown in Figure \ref{fig:UCTEfittings}. The methodology that has been used is that described in Ref.\cite{Clauset2007}, where a maximum likelihood approach is proposed to estimate the heavy tailed function from the data and a significance test is constructed for testing the plausibility of the power law. Measures shown in Table \ref{tab:UCTEfittings} are generic statistics on one side and results of the aforementioned statistical analysis on the other. We assume a quantity $x$ follows a power law if it is drawn from a probability distribution $p(x)\propto{x^{-\alpha}}$, where $\alpha$ is the scaling parameter of the distribution. Since the probability density of a power law distribution diverges as $x\rightarrow{0}$, there must exist a lower bound to the power law behavior \cite{NewmanPareto}. We denote this lower bound as $x_{min}$ and the number of events contained in the upper range as $n_{tail}$. Finally, the $p$-value denotes the significance test result: the power law is ruled out if $p\leq{0.1}$. As we can see, the power law model is a plausible one for every data set considered (i.e., the $p$-value for the best fit is sufficiently large) and  the scaling parameter values are similar to those encountered in the literature for the ENS and TLP distributions \cite{Dobson2007, Chen01}. Yet the power law model explains only a small amount of events: 15\% for ENS ($n_{tail}=104$), less than 10\% for TLP ($n_{tail}=57$) and 17\% for RT ($n_{tail}=157$) (even though it holds the better fit, with $p=0.73$). We believe that measures such as $n_{tail}$ and $x_{min}$  are fundamental to estimate the span of the power law behavior and to develop further quantitative models, yet these values have not been considered in any of the aforementioned references. Only in the reanalysis of Ref.\cite{Carreras2002} done in Ref.\cite{Clauset2007} we have found an estimate for $n_{tail}$ that gives an explanation for a 28\% of the events. 

We assume that the limited span of available data in each set might have a sensible influence in the final power law fitting outcome. It is nonetheless evident from these results that (a) power law behavior can not be assumed for the whole data observed, (b) we can not accept the existence of any critical point at this stage of the data span and (c) there must be right now considerably more dynamics not explained by the power law model.

\begin{table*}
\boldmath
\renewcommand{\tabcolsep}{2.1 mm}
\renewcommand{\arraystretch}{1.2} 
\begin{center} 
\begin{tabular}{ c| c c |c c| c c |c c|}
\cline{2-9}
 & \multicolumn{2}{c|}{\textbf{log-normal}} & \multicolumn{2}{c|}{\textbf{exponential}} & \multicolumn{2}{c|}{\textbf{Weibull}} & \multicolumn{2}{c|}{\textbf{PL + cut off}} \\ 
\cline{1-1}
\multicolumn{1}{|c|} {\textbf{Data set}} &  $LR$ & $p$ & $LR$ & $p$ & $LR$ & $p$ & $LR$ & $p$ \\
\cline{1-9}
\multicolumn{1}{|c|}{\textbf{ENS}} 	&  -0.405 & 0.68 & 2.64 & 0.00 & -0.393 & 0.69 & -0.419 &  0.36 \\ 
\multicolumn{1}{|c|}{\textbf{TLP}} 	&  0.319 & 0.75 & 3.42 & 0.00  & 0.467 & 0.64 &  -0.08 &  0.68 \\ 
\multicolumn{1}{|c|}{\textbf{RT}} 	&  -0.382 & 0.70 & 7.57 & 0.00  & -0.329	& 0.74 &  -1.7 &  0.06 \\ 
\cline{1-9}                       
\end{tabular}
\end{center}
\unboldmath
\caption{Test of power-law behavior. Positive values of the log likelihood ratios \textit{LR} favors the power law model. Values of $p\geq{0.1}$ imply though that result can not be trusted. The exponential distribution is definitely ruled out as possible model and only for the restoration time \textit{RT}, the power law with cut off could be considered a valid model. }
\label{tab:PowerLawTest}
\end{table*}

In order to check if other distributions may be a better fit, we have performed log likelihood and $p$-value tests with respect to log-normal, exponential, stretched exponential (Weibull) and power law with cut off distributions. Results are shown in  Table \ref{tab:PowerLawTest}. Positive log likelihood values favor the power law hypothesis and $p$-values higher than 0.1 imply no significance on the results. Log-normal, Weibull and exponential distributions can be ruled out as $p\geq{0.1}$ in the first two and $p=0$ in the latter with positive $LR$ values. Power law with cut off can be ruled out in the ENS and TLP data sets though it is a plausible option for the RT data set.


\section{Summary and discussion}

Power outages are considered unexpected phenomena in power grids. They appear without warning and, though widely investigated, there is not a common accepted theory that explains neither their pervasiveness nor their inner dynamics. The statistical overabundance of big blackouts has been explained using theories of systems failure able to reproduce their empirically found probability distribution. This distribution is considered a power law for most of the literature encountered, with the consequences that this algebraic tail involves (i.e., self-organization, criticality and universalities). In order to add one more reference to this field, in this paper we have analyzed seven years of disturbances data for the UCTE power grid and for three major event measures: energy not supplied, total loss of power and restoration time. Although evidences for self-organized criticality have been suggested for even five years of data \cite{Carreras00}, support for the power law hypothesis has been found moderate for two (ENS and TLP) of the three measures considered. Moreover, the amount of events explained by the power law hypothesis can be considered negligible. This facts make it difficult to accept the existence of an equilibrium point near criticality for the UCTE power grid, at least at this stage of data analysis, and it also suggests that most of the power grid dynamics should be explained by different models, other than power law.

There still exist many complexities not explained in this systems. Ongoing research is now focused in analyzing major events probability distributions in connection with (a) the reasons that trigger these major events and (b) the structure and topology of the power grids involved in these major malfunctions. Results will be published elsewhere.


\begin{acknowledgments}

We deeply thank David Chury, Data and Modeling Expert at the UCTE, for 
his patience, kindness and useful discussions. We also thank Aaron Clauset
for enlightening comments and clarifications about distributions with heavy tails.
One author (MRC) wishes to thank specially Selina Delgado-Raack for most helpful conversations and support.
This work has been supported by the Santa Fe Institute and the Technical University of Catalonia, UPC.

\end{acknowledgments}



\end{document}